\documentclass{cpcauth}

\newcommand{\varql}{{\mathrm{Var}(\ql)}}
\newcommand{\ql}{q_{l}}
\newcommand{\absq}{\left|q\right|}
\newcommand{\ml}{\mu_{l}}

\usepackage{epsfig}
\usepackage{amssymb}
\usepackage{goodfloat}

\begin{document}

\begin{frontmatter}

\title{Monte Carlo Simulations of Vector Spin Glasses at Low Temperatures}
\author{Helmut G. Katzgraber}

\address{Department of Physics, University of California, Santa Cruz, CA 95064}

\begin{abstract}
In this paper I report results for simulations of the three-dimensional gauge glass
and the four-dimensional $XY$ spin glass using the parallel tempering 
Monte Carlo method at low temperatures for moderate sizes.
The results are qualitatively consistent with earlier work on the 
three- and four-dimensional Edwards-Anderson Ising spin glass. 
I find evidence that large-scale excitations may cost only a finite amount of 
energy in the thermodynamic limit. The surface of these excitations is 
fractal, but I cannot rule out for the $XY$ spin glass a scenario
compatible with replica symmetry breaking where the surface of 
low-energy large-scale excitations is space filling.
\end{abstract}

\end{frontmatter}

There has been an ongoing controversy regarding the spin glass phase.
There are two main theories: the ``droplet picture''
(DP) by Fisher and Huse \cite{fisher:87} and the replica symmetry breaking
picture (RSB) by Parisi \cite{parisi:79}. 
While RSB follows the exact solution of the Sherrington-Kirkpatrick model
and predicts that excitations involving a finite fraction of the spins
cost a finite energy in the thermodynamic limit, the droplet picture states 
that a cluster of spins of size $l$ costs an energy proportional to
$l^{\theta}$ 
where $\theta > 0$. It follows that in the thermodynamic limit,
excitations that flip a finite cluster of spins cost an infinite energy.
In addition, the DP states these excitations are fractal with a fractal 
dimension $d_s < d$, where $d$ is
the space dimension, whereas in RSB these excitations
are space filling, i.e. $d_s = d$.
Krzakala and Martin, as well as Palassini and Young 
\cite{kmpy} (referred to as KMPY) find, on the
basis of numerical results on small systems with Ising symmetry,
that an intermediate picture may be present: while
the surface of large-scale excitations appears to be fractal, only a finite 
amount of energy is needed to excite them in the thermodynamic limit. 

The differences between DP and RSB can be quantified by studying the distributions
$P(q)$ and $P(\ql)$ of the spin overlap $q$ and link overlap $\ql$.
For finite systems, the DP predicts two peaks at $\pm q_{\rm EA}$, 
where $q_{\rm EA}$ is the
Edwards-Anderson order parameter, as well as a tail down to $q = 0$ that
vanishes in the thermodynamic limit, like $\sim L^{-\theta}$ for perturbations
introduced by a change in boundary conditions\footnote{In this work I use
$\theta^\prime$ instead of $\theta$ as I introduce the excitations thermally
with {\em fixed} boundary conditions \cite{katzgraber:01}.}.
In contrast, RSB predicts a non-trivial distribution with a finite weight
in the tail down to $q = 0$, independent of system size.
In addition, DP predicts the variance of the link
overlap to fit a power law of the form
\begin{equation}
{\rm Var}(\ql)  = a + bL^{-\mu_l}
\label{2params}
\end{equation}
where $a = 0$ and, as shown in
Ref.~\cite{katzgraber:01}, $\mu_l = \theta^\prime +2(d - d_s)$, whereas for RSB 
one expects $a > 0$.

While there has been considerable work on Ising-type spin glass
systems, only few attempts \cite{katzgraber:01a,katzgraber:01c} to understand 
the nature of the spin glass phase for models with a vector order parameter
have been made.
In this work I review recent results for the three-dimensional (3D) gauge glass
and the four-dimensional (4D) $XY$ spin glass.

The Hamiltonian of the vector models analyzed can be summarized by 
\begin{equation}
{\mathcal H} = - \sum_{\langle i, j\rangle} J_{ij }
\cos(\phi_i - \phi_j - A_{ij}),
\label{hamiltonian}
\end{equation}
where the sum ranges over nearest neighbors on a hypercubic lattice in $d$ 
dimensions of size $N = L^d$. Here $\phi_i$ represent the angles of the $XY$ spins
as ${\bf S} = (\cos(\phi),\sin(\phi))$ and therefore $|{\bf S}| = 1$. 
Periodic boundary conditions are applied.
In the case of the gauge glass, I set $d = 3$ and $J_{ij} = 1$ for all $i, j$.
The $A_{ij}$ are quenched random variables uniformly distributed between 
$[0,2\pi]$ representing the line integral of the vector potential        
directed from site $i$ to site $j$.
For the 4D $XY$ spin glass I set $d = 4$ and $A_{ij} = 0$
for all $i,j$. The $J_{ij}$ are chosen according to a Gaussian distribution 
with zero mean and standard deviation unity.

The spin overlap $q$ is traditionally defined as
\begin{equation}
q = \frac{1}{N}\sum_{i=1}^N {\bf S}_i^{\alpha} \cdot {\bf S}_i^{\beta} \; ,
\end{equation}
where $\alpha$ and $\beta$ represent two replicas of the system with the same
disorder. For this quantity
to be a sensible order parameter, it has to be maximized with respect to all
symmetries of the Hamiltonian. This is described in more detail in 
Refs.~\cite{katzgraber:01a} and \cite{katzgraber:01c} for the gauge glass and
$XY$ spin glass, respectively. In addition I introduce the link overlap $\ql$
defined by
\begin{equation}
\ql = \frac{1}{N_b}\sum_{\langle i,j\rangle}
({\bf S}_i^{\alpha}\cdot{\bf S}_j^{\alpha})
({\bf S}_i^{\beta}\cdot{\bf S}_j^{\beta}) \; ,
\end{equation}
with $N_b = dN$ the number of bonds.

To avoid critical effects influencing the data, I perform the simulations at low
temperatures, typically $T \le 0.2T_c$. To equilibrate
the systems at such low temperatures, I use the parallel tempering Monte Carlo
method \cite{pt}.
For equilibration tests for parallel
tempering Monte Carlo on the models studied
I refer the reader to Refs. \cite{katzgraber:01a} and
\cite{katzgraber:01c}.

\section*{Results}
\label{results}

\begin{figure}
\centerline{\epsfxsize=6.5cm \epsfbox{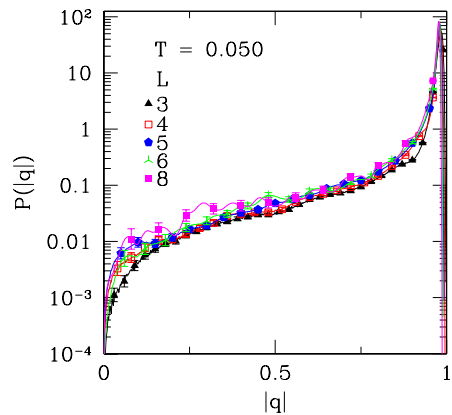}}
\vspace{-0.6cm}
\caption{ Data for the overlap distribution
$P(\absq)$ at temperature $T = 0.050$
for the 3D gauge glass. Note the logarithmic vertical scale.
In this and other similar figures in the paper, I only display a subset of
all the data points  while the lines connect all the data points in the set.
Thus, the structure in the lines between neighboring symbols is meaningful.
}
\label{pq0.050-gg}
\end{figure}

\begin{figure}
\centerline{\epsfxsize=6.5cm \epsfbox{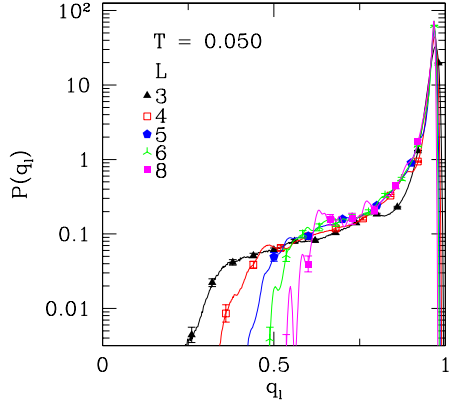}}
\vspace{-0.6cm}
\caption{ The distribution of the link overlap at $T = 0.050$ for the 3D gauge
glass. Note the logarithmic vertical scale.}
\label{pqb0.050-gg}
\end{figure}

Figures \ref{pq0.050-gg} and \ref{pqb0.050-gg} show data\footnote{I present
data for $\absq$ since for the gauge glass $q \in {\mathbb
C}$.} for $P(\absq)$ and $P(\ql)$ at $T = 0.050$, 
well below $T_c \approx 0.45$ \cite{olson:00} for the 3D gauge glass. 

There is a clear peak in $P(\absq)$ for large $\absq$ as well as a tail 
at small $\absq$. The weight in the tail does not decrease with increasing
$L$, as is expected in the standard interpretations of the droplet
theory. If anything, the weight increases for larger sizes. 
There is some evidence that $P(\absq)$ 
stays flat down to smaller $\absq$ for larger $L$, although the range of
sizes is too small to make a reliable extrapolation. 

As with the distribution of $\absq$, there is a pronounced peak at large
$\ql$-values in the distribution of $P(\ql)$ for the 3D gauge glass. 
Note also the appearance 
of a smaller peak at smaller $\ql$ for larger system sizes, as predicted by
RSB. 

\begin{figure}
\centerline{\epsfxsize=6.5cm \epsfbox{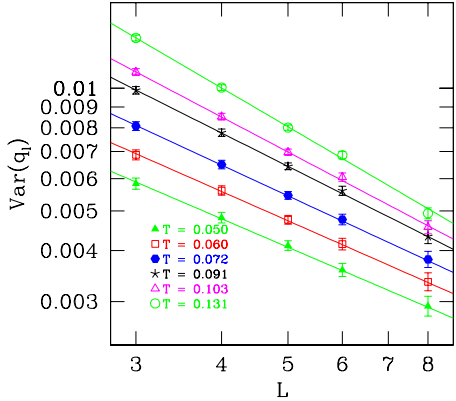}}
\vspace{-0.6cm}
\caption{Log-log plot of the variance of $\ql$ vs.~$L$ at several
temperatures for the 3D gauge glass.
}
\label{varqb-gg}
\end{figure}

The width of the distribution decreases with increasing system size.
This can be seen in Figure \ref{varqb-gg} where I plot the variance of the
link overlap as a function of system size $L$. The data is consistent with a
power law decrease ($a = 0$ in Eq.~\ref{2params})
where the (presumably effective) exponent
varies slightly with $T$. Extrapolating to $T = 0$ gives 
$\ml \equiv \theta^\prime + 2(d - d_s) = 0.501 \pm 0.04$.
Assuming $\theta^\prime \approx 0$ I find $d - d_s = 0.25 \pm 0.02$,
implying that for the 3D gauge glass system-size excitations have a fractal 
surface in the thermodynamic limit as predicted by the droplet picture.

\begin{figure}
\centerline{\epsfxsize=6.5cm \epsfbox{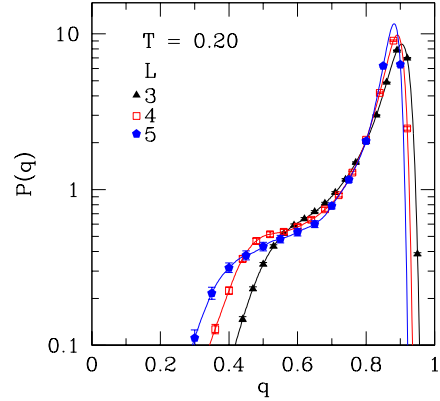}}
\vspace{-0.5cm}
\caption{
Data for the overlap distribution $P(q)$ at $T = 0.20$ 
for 4D $XY$ spin glass. Note the logarithmic vertical scale. 
}
\label{pqd-0.20-xy}
\end{figure}

\begin{figure}
\centerline{\epsfxsize=6.5cm \epsfbox{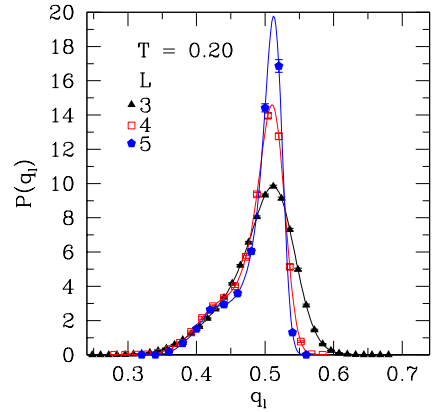}}
\vspace{-0.5cm}
\caption{
The distribution of the link overlap at $T = 0.20$ for the 4D $XY$ spin glass.
}
\label{pqb-0.20-xy}
\end{figure}

Figures \ref{pqd-0.20-xy} and \ref{pqb-0.20-xy} show data for $P(q)$ 
and $P(\ql)$ for $T = 0.20$ (to be compared with $T_c \approx 0.95$ 
\cite{jain:96}) for the 4D $XY$ spin glass. Again one sees a large peak for large
$q$ and $\ql$ values. The data for $P(\ql)$ exhibits a hint of a shoulder for
smaller values of $\ql$.

\begin{figure}
\centerline{\epsfxsize=6.5cm \epsfbox{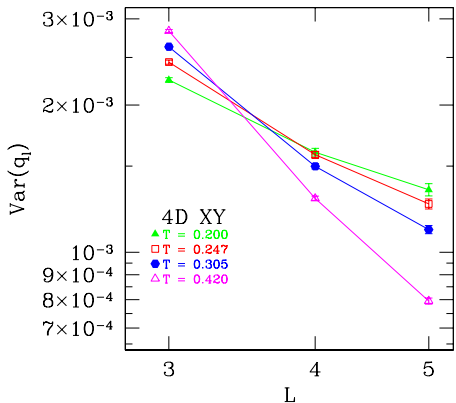}}
\vspace{-0.5cm}
\caption{
Log-log plot of the variance of $\ql$ vs.~$L$ at several 
temperatures for the 4D $XY$ spin glass.
}
\label{varqb-xy}
\end{figure}

The width of
the distribution decreases with increasing system size. This is demonstrated
in Fig.~\ref{varqb-xy} where I show the variance of $\ql$
against system size $L$ for several low temperatures.
There is some curvature in the data for $\varql$ so first I attempt a
three-parameter fit to Eq.~\ref{2params}. As there are the same number of data points
as variables, I cannot assign fitting probabilities to the fits but from the
data I find $a = 0.00100$, $0.00087$, $0.00073$ and $0.00036$ for $T = 0.200$, 
$0.247$, $0.305$ and $0.420$, respectively. 

I also attempt a power law fit to Eq.~\ref{2params} setting $a = 0$.
The quality of the fits is poor with 
probabilities $Q = 5.0 \times 10^{-2}$, $3.6 \times 10^{-3}$,
$2.9 \times 10^{-6}$ and $6.0 \times 10^{-8}$ for $T = 0.200$, $0.247$, 
$0.305$ and $0.42$, respectively\footnote{The complete set of fitting
parameters for both fitting functions presented
in Eq.~\ref{2params} can be found in
Ref.~\cite{katzgraber:01c}.}. The effective
exponent $\ml$ is found to vary with temperature.
Extrapolating to $T = 0$, I obtain $\ml \equiv \theta^\prime
+ 2(d - d_s) =  0.294 \pm 0.073$.
Assuming $\theta^\prime \approx 0$, one obtains
$d - d_s = 0.147 \pm 0.036$.

\section*{Conclusions}
\label{conclusions}

To conclude, I have studied the properties of the 3D gauge 
glass and the 4D $XY$ spin glass at low temperatures. 
For both models, the order parameter distribution $P(q)$ has, in addition to a
peak, a tail that seems to extend for smaller values of $q$ and
whose height seems to persist as the system size increases. This
interpretation of the data is compatible with the RSB picture or the KMPY 
scenario. The range of lattice sizes is very small, however, so this 
conclusion can be considered at most tentative.

For the gauge glass, the variance of the link overlap indicates that the
surface of low-energy large-scale excitations is fractal with 
$d - d_s = 0.25 \pm 0.02$ in agreement with the KMPY scenario.
The log-log plot for $\varql$ for the $XY$ spin glass shows curvature, 
possibly indicating
a non-zero value in the thermodynamic limit, a result compatible with
RSB. Due to the small range of system sizes, however, I cannot rule 
out the possibility that $\varql \to 0$ at large $L$, which is compatible 
with the KMPY scenario or droplet picture.
The effects of vortices in the spin glass phase still remain to
be understood. It would be useful to look more carefully at the nature of the
large-scale low-energy excitations to see whether they correspond to gradual
orientations in the spin directions or whether vortices play a role.\newline

I would like to thank A.~P.~Young for useful discussions.
This work was supported by the National Science Foundation under grant
DMR 0086287 and a Campus Laboratory Collaboration (CLC).

\end{document}